\documentclass{aa}
\usepackage{graphics}
\begin{document}
\title{Lithium abundances from the 6104{\AA} line in cool Pleiades stars}

\author{A. Ford
\inst{1,2}
\and R.D. Jeffries
\inst{1}
\and B. Smalley
\inst{1}}
\offprints{A. Ford}
\mail{alison.ford@open.ac.uk}
\institute{Department of Physics, Keele University, Keele, Staffordshire, ST5 5BG. UK 
\and Department of Physics and Astronomy,  The Open University, Walton Hall, Milton Keynes, MK7 6AA, UK\\
email: alison.ford@open.ac.uk or rdj@astro.keele.ac.uk or bs@astro.keele.ac.uk}
\date{Received; accepted}
\authorrunning{A. Ford et al.}

\abstract
{Lithium abundances determined by spectral synthesis from both the
6708{\AA} resonance line and the 6104{\AA} subordinate line are
reported for 11 Pleiades late-G and early-K stars observed at the
William Herschel Telescope. Firm detections of the weak subordinate
line are found for four objects, marginal detections for four, and
upper limits for the remaining three stars.  Some of these spectra were previously
analysed by Russell (1996), where he reported that abundances derived
from the 6104{\AA} line were systematically higher than those obtained from the
6708{\AA} line by 0.2-0.7\,dex. He also reported a reduced spread
in the 6104{\AA} line abundances compared with those determined from
the 6708{\AA} feature. Using spectral synthesis we have re-analysed Russell's data, along with our own. Our results do not entirely support
Russell's conclusions. We report a $\sim$0.7\,dex scatter in the
abundances from 6708\AA\ and a scatter at least as large from the
6104\AA\ line. We find that this is partly explained by our inclusion
of a nearby \ion{Fe}{ii} line and careful modelling of damping wings in
the strong metal lines close to the 6104\AA\ feature; neglect of these
leads to overestimates of the Li abundance which are most severe in
those objects with the weakest 6104\AA\ lines, thus reducing the
abundance scatter.  We find a reasonable correlation between the
6104{\AA} and 6708{\AA} Li abundances, although four stars have
6104{\AA}-determined abundances which are significantly larger than the
6708{\AA}-determined values by up to 0.5\,dex, suggesting problems with
the homogeneous, 1-dimensional atmospheres being used. We show that
these discrepancies can be explained, although probably not uniquely,
by the presence of star spots with plausible coverage fractions. The
addition of spots does not significantly reduce the apparent scatter in
Li abundances, leaving open the possibility that at least some of the
spread is caused by real star-to-star differences in pre-main sequence
Li depletion.
}

\titlerunning{\ion{Li}{i} 6104\AA\ in the Pleiades}

\maketitle

\section{Introduction}
\label{sec-intro}
The observed spread of lithium abundances in stars of the young
($\sim$100 Myr) zero-age main sequence (ZAMS) \object{Pleiades} cluster has been
puzzling observers for around 20 years. The spread appears at
$T_{\mathrm{eff}}$ $\sim$ 5800\,K and continues into cooler stars. The
magnitude of the spread increases with decreasing mass and reaches a
maximum (more than 1\,dex) at 0.8M$_{\odot}$ in the mid-K stars before
possibly decreasing in the coolest objects (Jones et al. \cite{jones96}). Duncan
\& Jones (\cite{dj83}) first reported the scatter, and interpreted it
as being due to a large ($\sim$0.4~Gyr) age spread within the cluster
stars, although this has since been shown to be unlikely (Soderblom et
al. \cite{s93a}). A scatter is also seen in other young clusters
(e.g. Randich \cite{randich01}), but
diminishes by the age of the Hyades (Soderblom et al. \cite{s95} --
although most Li data for Hyades K stars are upper limit
estimates). Pasquini et al. (\cite{pasquini97}) report evidence that
the spread reappears with a dispersion of~$>$1\,dex among the solar-type
stars of the old ($\sim$5 Gyr) \object{M67} cluster.

Soderblom et al. (\cite{s93a}) noted that there is a connection between
rotation and Li abundance in these cool stars, in the sense that rapid
rotators seem to display higher Li abundances. Jeffries (\cite{j2000})
showed that {\em slow} rotators can have {\em either} high {\em or} low
abundances. In addition, there is a correlation with chromospheric
activity, with more-active stars also showing more lithium. It is not
clear whether this link is causal or coincidental: {\em i.e.} whether
star spots or other chromospheric effects give rise to an {\em
apparent} abundance spread (Stuik et al. \cite{stuik97};
King et al. \cite{king2000}), 
or if the spread is real and linked to
activity through a physical mechanism connected with the rotation in
these objects (Soderblom et al. \cite{s93a}; Jones et
al. \cite{jones97}). For instance, non-standard effects such as
rotation-driven mixing (e.g. Chaboyer et al. \cite{chaboyer95}) or perhaps even metallicity differences between
individual cluster stars could cause variations in star-to-star lithium
depletion during the PMS phase.

Almost all Li abundance measurements in main sequence Pop. I and Pop. II stars
have been made using the strong \ion{Li}{i}~6708{\AA} resonance line.
This line is formed high in the atmosphere and samples a limited range
of depths. It is plausible that it could be affected by chromospheric
activity (Houdebeine \& Doyle \cite{hd95}) or other inhomogeneities
such as star spots (Giampapa \cite{gia84}), leading to erroneous Li
abundances. Stuik et al. (\cite{stuik97}) argue that the effects of the
chromosphere are primarily communicated through photospheric
stratification and ionizing radiation fields, and therefore by
changes in the ionization balance. To first order, all the alkali lines
should be similarly affected by the presence of a chromosphere. Indeed,
a number of authors have noted that the \ion{K}{i} line at 7699\AA\
also shows some evidence for a spread in strength at the same effective
temperatures, although perhaps not as large as for \ion{Li}{i}~6708\AA\
(Soderblom et al. \cite{s93a}; Jeffries \cite{j99a}; King et
al. \cite{king2000}).

Explaining the observed spread in Li abundances among such a group of
co-eval stars is an important goal. Either the abundance spread is
real, which would tell us that non-standard mixing processes
(those additional to convection) can produce star-to-star differences,
probably as a result of differing angular momenta, or there
really is no abundance spread; in which case the observations would
tell us that crude, one-component, plane-parallel atmospheres poorly
describe conditions in young, convective stars where there might be
additional turbulence or atmospheric inhomogeneities to contend with.

A crucial test of the reliability of the atmospheres used in Li
abundance analyses would be to obtain the abundance using alternative
lines. Unfortunately the available optical lines (the subordinate transitions at
6104{\AA} and 8126{\AA}) are very weak and blended with other strong
metal features. Nevertheless, these lines sample different depths
within the atmosphere and the lower levels of their transitions form
the upper level for the resonance line. It is, therefore, of primary
importance to determine Li abundances from these lines separately, and
to compare them with abundances measured from the resonance line.

Russell (\cite{russell96}) measured 6104 (hereafter `6708' and `6104'
are taken to mean `the \ion{Li}{i} resonance line at 6708{\AA}' and
`the \ion{Li}{i} subordinate line at 6104{\AA}' respectively) in the
atmospheres of six Pleiades late-G/early-K stars, and compared the abundances to
those measured from 6708. His results suggested that 6104
gave abundances which were systematically higher than those from 6708
by 0.2-0.7\,dex. He also reported a significantly smaller Li
abundance spread derived from 6104. If Russell's conclusions were correct, they
would indicate that something is seriously wrong with the model
atmospheres used in such abundance determinations. The combination of a
reduction in spread from one line over the other, and an abundance
discrepancy between the two lines would suggest that there are problems
with the temperature structure of the atmospheres, possibly related to
inhomogeneities such as spots or plages on the stellar
surface. However, Mart\'{\i}n (\cite{martin97}) claims that Russell has
probably overestimated the strength of the subordinate line by a factor
of 2 or 3, due to his use of an inappropriate Gaussian-fitting technique. If
this is taken into account, Mart\'{\i}n believes that the abundances
from the two lines might agree to within 0.2\,dex; in this case the
agreement between the lines would support the validity of the
atmospheric analysis, confirm the abundance spread, and suggest
that non-standard mixing in PMS stars is important.

Given this controversy, we have repeated Russell's experiment,
collecting our own data for several Pleiades G and K stars. We also
obtained the data (including calibration frames) taken by Russell in
1993. These data were extracted, calibrated and analysed in exactly the
same way as our own.

\section{Observations and Reduction}

A sample of 6 late-G and early-K stars were selected from the catalogue
of Soderblom et al. (\cite{s93a}) on the basis of $V$ (11.0 to 12.0) and
$B-V$ (0.75 to 0.9) colours, and low $v \sin i$
($<$10~km~s$^{-1}$). These objects show a large spread in A(Li) (=
$\log_{10} \left(\frac{N({\mathrm{Li}})}{N({\mathrm{H}})}\right)$ + 12) at a given
temperature, but have relatively low projected rotational velocities,
simplifying the deblending and measurement of the lines. The objects
have previously reported A(Li) values between 1.59 and 3.19
(Soderblom et al. \cite{s93a}). Two F stars were also observed for
comparison, but their $v \sin i$ values were too large to attempt to
determine abundances from the severely blended 6104\AA\ line (see
below). The new spectroscopic data acquired for this study were taken
on the 4.2m William Herschel Telescope (WHT) on 1998 November 28 and
29, and 1999 December 22 and 23. The Utrecht Echelle Spectrograph (UES)
was used for its high resolving power ($\sim$50000), with the E31 (31
grooves mm$^{-1}$) grating. Typical exposure times of 6000-9000 seconds were
used, split over several exposures in order to minimise cosmic-ray
contamination. The central wavelength was chosen to give good
coverage of both Li lines.  Russell's data were retrieved from the ING
Data Archive (maintained by the Cambridge Astronomical Survey
Unit). These data were obtained on 1993 January 8-10, using the same
spectrograph and gratings, but a smaller EEV CCD and slightly shorter
exposure times.  Observational
parameters are summarised for all datasets in Table~\ref{table-obsparam}.
Target information is summarised in Table~\ref{table-data1}.

\begin{table*}
\caption[Observational information for data contained in this chapter,
and for Russell (1996)]{Observational information for new data obtained
for this paper and for data from Russell (1996).}
\begin{center}
\begin{tabular}{llll}
\hline
 & This paper & & Russell (1996) \\
\hline
Observation date    & 1998 11 28-29 & 1999 12 22-23 & 1993 01 08-10 \\
Echelle grating     & E31           & E31           & E31 \\
Detector            & SITe1          & SITe1        & EEV6 \\
Pixel size          & 24.0 $\mu$m  & 24.0$\mu$m     & 22.5$\mu$m  \\
Dispersion at 6104{\AA} & 0.061{\AA} pixel$^{-1}$ & 0.061{\AA} pixel$^{-1}$ & 0.057{\AA} pixel$^{-1}$ \\ 
Central wavelength  & 5875{\AA}     & 5814{\AA}     & 6770{\AA} \\
Exposure time       & 9000 s (5 $\times$ 1800 s)   & 6000 s (5
 $\times$ 1200 s) & 3600 s (3 $\times$ 1200 s) \\
Resolving power & 50\,000   & 40\,000   & 48\,000 \\
\hline
\end{tabular}
\end{center}
\label{table-obsparam}
\end{table*}

The usual calibration frames were taken including tungsten flat fields,
bias frames and thorium-argon arc spectra. We also obtained a solar
spectrum. The
data were extracted and wavelength calibrated using the Starlink {\sc
echomop} (Mills et al. \cite{mills97}) and {\sc figaro} (Shortridge et
al. \cite{shortridge99}) packages. Scattered-light subtraction was
performed during the {\sc echomop} reduction, using inter-order
regions. Continuum placement was determined by fitting a polynomial to
line-free regions of the spectrum around the Li lines, as determined by
a spectrum synthesis at the appropriate temperature and
metallicity. The $rms$ of this fit was used to determine a conservative
signal-to-noise (S/N) estimate, such that $\frac{1}{rms}$ = S/N (values are presented in
Table~\ref{table-data1}). The theoretical S/N values, based simply on the
number of detected photons, should be $\sim$30\% larger: the discrepancy probably 
arises from weak or unidentified lines, residual cosmic ray features or
flat-fielding errors. Russell claimed S/N values of 100-180 for his
observations, but these values are significantly larger than either the
conservative S/N estimates or the theoretical S/N levels that we
determine.

\begin{figure*}
\resizebox{\hsize}{!}{\includegraphics{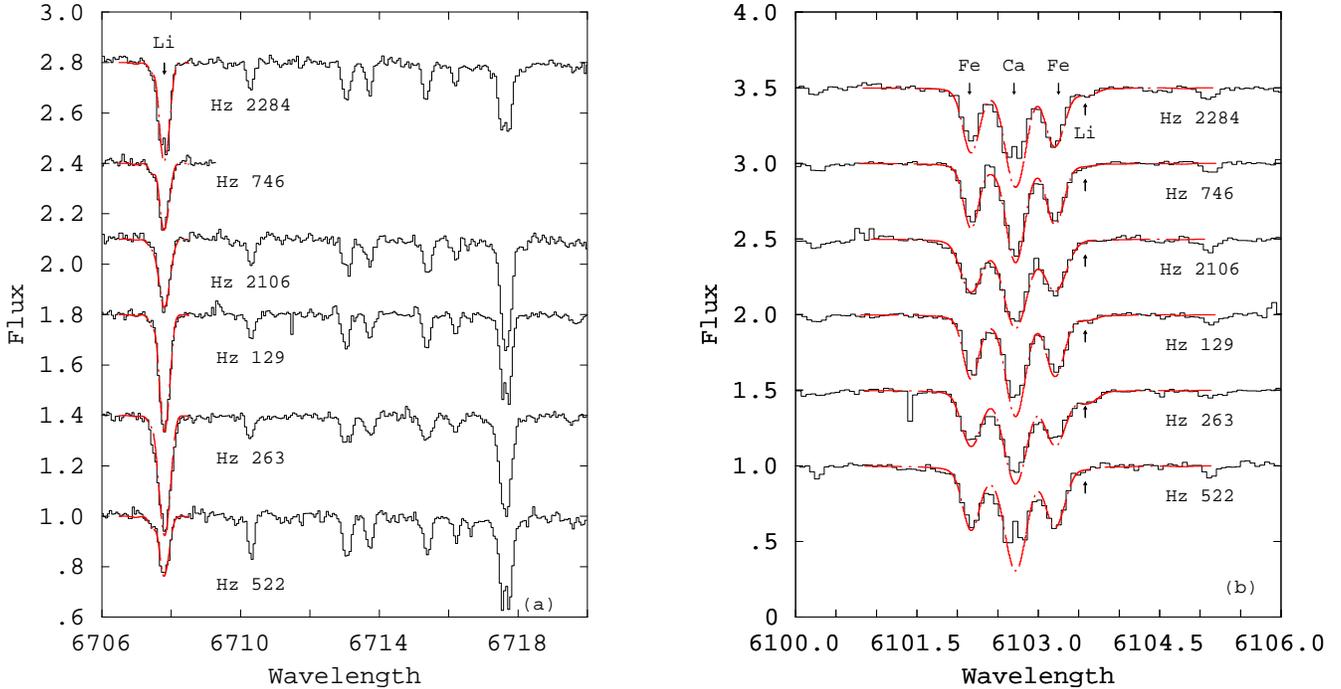}}
\caption{Normalised spectra for our sample stars in the region of {\bf (a)} 6708 and {\bf (b)} 6104. Solid lines represent the observed data, while dashed lines represent the synthetic fits. In objects where the 6104 line was not detected, the syntheses are for A(Li)~=~0.}
\label{fig-spec74}
\end{figure*}

\begin{figure*}
\resizebox{\hsize}{!}{\includegraphics{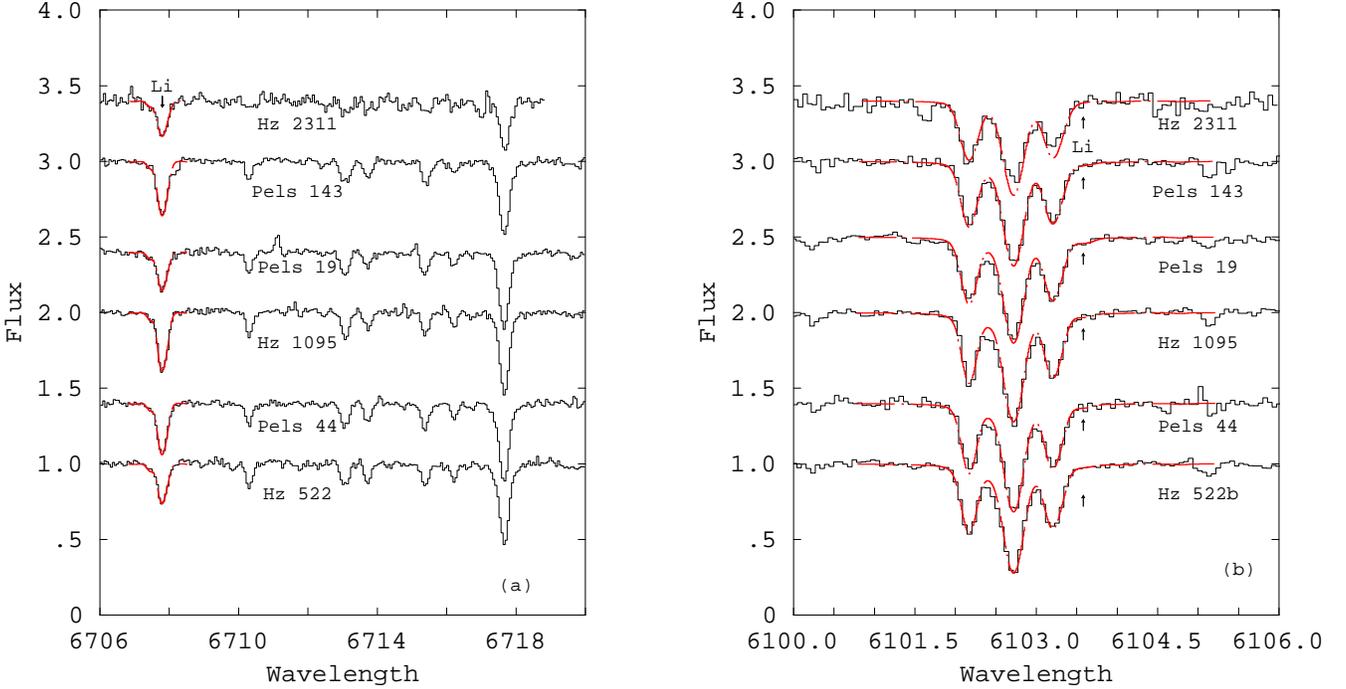}}
\caption{Normalised spectra for Russell's data in the regions of the {\bf (a)}
6708 and {\bf(b)} 6104 lines (solid line) and fitted synthetic LTE spectra (dashed
line). Note that the spectra have been offset for clarity. Arrows
indicate the wavelengths of the Li features}
\label{fig-russspec}
\end{figure*}

Fig.~\ref{fig-spec74} shows reduced spectra in the regions of 6708 and
6104 for our data. Spectra for those objects observed by Russell are
presented in Fig.~\ref{fig-russspec}.

\begin{table*}
\caption[Object photometry (from Soderblom et al. 1993a) and model
parameters]{Object photometry (from Soderblom et al. 1993a), model
parameters, S/N and resolution of spectra. Objects below the line are those in Russell's sample.}
\begin{center}
\begin{tabular}{llllllllll}
\hline
Object   & $B-V$ & $T_{\mathrm{eff}}$ & $v\sin i$   & $\xi$       & S/N    & S/N    & FWHM   & FWHM   \\
         &       & (K)                &~km~s$^{-1}$ &~km~s$^{-1}$ & (6708) & (6104) & (6708) & (6104) \\
\hline
\object{Hz 1122}  & 0.425 & 6535               & 28.6$\pm$1.2 & 2.40    & 135    & 135    & 0.17\AA  & 0.16\AA \\ 
\object{Hz 233}   & 0.493 & 6290               & 14.1$\pm$0.7 & 1.80    & 140    & 115    & 0.17\AA  & 0.16\AA \\ 
\object{Hz 2284}  & 0.743 & 5449               &  3.7$\pm$0.8 & 2.88    &  90    &  95    & 0.17\AA  & 0.16\AA \\ 
\object{Hz 746}   & 0.768 & 5383               &  4.8$\pm$0.8 & 2.10    &  75    &  90    & 0.14\AA  & 0.12\AA \\ 
\object{Hz 2106}  & 0.823 & 5245               &  8.0$\pm$0.8 & 2.32    &  80    &  70    & 0.17\AA  & 0.16\AA \\ 
\object{Hz 129}   & 0.830 & 5228               &  5.3$\pm$1.4 & 2.02    &  95    & 110    & 0.14\AA  & 0.12\AA \\ 
\object{Hz 263}   & 0.836 & 5213               &  7.8$\pm$0.8 & 2.13    & 100    &  90    & 0.14\AA  & 0.12\AA \\ 
\object{Hz 522}   & 0.879 & 5110               &  4.4$\pm$0.8 & 2.34    &  80    &  80    & 0.17\AA  & 0.16\AA \\ 
\hline

\object{Hz 2311}  & 0.780 & 5352               &  6.5$\pm$0.8 & 2.3     &  40    &  30    & 0.16\AA  & 0.13\AA \\
\object{Pels 143} & 0.853 & 5177               &  5.2$\pm$1.2 & 2.3     &  50    &  45    & 0.16\AA  & 0.13\AA \\
\object{Pels 19}  & 0.853 & 5172               &  4.8$\pm$1.4 & 2.3     &  75    &  65    & 0.16\AA  & 0.13\AA \\
\object{Hz 1095}  & 0.858 & 5160               &  3.6$\pm$0.8 & 2.3     &  75    &  90    & 0.16\AA  & 0.13\AA \\
\object{Pels 44}  & 0.855 & 5167               &  3.9$\pm$3.1 & 2.3     &  65    &  50    & 0.16\AA  & 0.13\AA \\
Hz 522  & 0.879 & 5110                         &  4.4$\pm$0.8 & 2.34    &  65    &  70    & 0.16\AA  & 0.13\AA \\
\hline
\end{tabular}
\end{center}
\label{table-data1}
\end{table*}

\section{Analysis}
\label{sec-analysis}
\begin{figure}
\resizebox{\hsize}{!}{\includegraphics{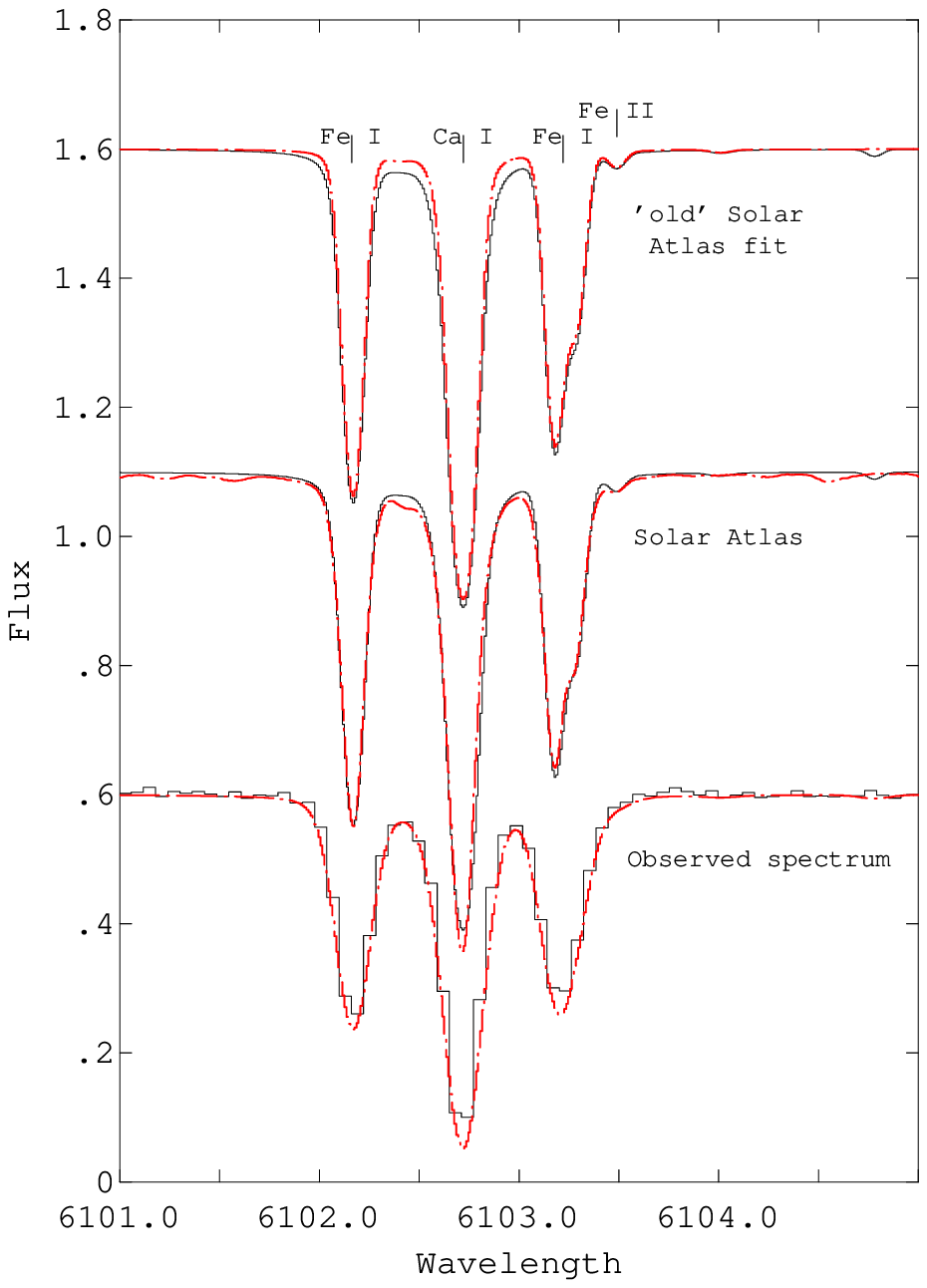}}
\caption{Fits (dashed lines) to the solar atlas (top and middle) and observed solar spectrum (bottom) in the region of 6104, using $gf$ factors determined for the lines. The top spectrum shows the synthesis using standard Van der Waals broadening co-efficients, while the ABO values were used for the other two syntheses (Sect.~\ref{sec-analysis}).}
\label{fig-solar}
\end{figure}

Effective temperatures were calculated from the $B-V$ photometry presented in
Soderblom et al. (\cite{s93a}) using the calibration of B\"ohm-Vitense
(\cite{bv81}). The surface gravity was taken as 4.5 for all stars in
the sample. The photometry, calculated $T_{\mathrm{eff}}$, $v \sin i$
(taken from Queloz et al. ~\cite{queloz98}), microturbulent velocity
($\xi$ -- see below), S/N, and instrumental broadening (FWHM --
obtained from Gaussian fits to arc lines) values are presented in
Table~\ref{table-data1}.  The models used in the analysis were Kurucz, 1-D,
homogeneous, LTE, {\sc atlas9} model atmospheres (Kurucz
\cite{kurucz93}) incorporating the mixing-length treatment of
convection ($\alpha$=1.25) without overshooting (Castelli et
al. \cite{castelli97}). Before any analysis of the object spectra was
undertaken, we used a high-resolution solar atlas (Kurucz et
al. \cite{kurucz84}), and solar spectra taken during our observing
time, to tune the atomic parameters of the lines adjacent to Li. This was done by
fitting a spectral synthesis to the data, assuming the solar parameters
to be: $T_{\mathrm{eff}}$~=~5777\,K; $\log g$~=~4.44;
$\xi$~=~1.5~km~s$^{-1}$; $v \sin i$~=~2~km~s$^{-1}$; A(Fe)~=~7.54;
A(Ca)~=~6.36. This assumed value for the microturbulence is a few
tenths higher than frequently used for the Sun, but is the value
espoused by Castelli et al. (\cite{castelli97}) in their examination of the {\sc atlas9}
model atmospheres. We will examine the effects of this assumption subsequently.
The complete line lists used for the 6708 and 6104
regions are presented in Tables~\ref{table-7lines} and~\ref{table-4lines} (it
should be noted that whilst all the lines listed were put into the
syntheses, many had negligible equivalent width -- hereafter
EW). Damping constants were those provided by the Kurucz \& Bell
(\cite{kb95}) line list, except for the \ion{Ca}{i} lines adjacent to 6104, and for the Li lines themselves,
where the larger Anstree, Barklem O'Mara (ABO -- Barklem et al. \cite{barklem00})
Van der Waals broadening parameters were used. We also arbitrarily 
increased the damping constants for the strong \ion{Fe}{i} lines (for
which no ABO results are yet available) and as a result, 
drastically improved the fits to the
line wings in the high resolution solar atlas (see Fig.~\ref{fig-solar}). 
The $gf$ values were varied for each line in turn, until
the minimum chi-squared value for each line was obtained. The resultant
line list and synthesis was broadened to the instrumental resolution, then checked against our observed solar spectrum. It was then used throughout the rest of the analysis.
We note that if we had assumed a smaller value for the solar
microturbulence then we would have simply increased the damping widths
by a little more to compensate.

$gf$ values were also obtained for $\sim$ 20 \ion{Fe}{i} lines in the
ranges 6000-6150{\AA} and 6700-6750{\AA} (listed in
Table~\ref{table-felines}, again by fitting the observed solar spectrum. The
EWs of the lines (obtained for each object by direct integration below
the apparent continuum) were then used to determine the microturbulence
parameter for each Pleiades target in our sample. The most likely $\xi$ value is
taken to be that where there is no correlation between EW and abundance
for the lines in the sample (Magain ~\cite{magain86}). The lines
measured had EWs between 0.01{\AA} and 0.16{\AA}. We find $\xi$ values
(listed in Table~\ref{table-data1}) of 2-3~km~s$^{-1}$ for the Pleiades targets
and estimate that the maximum likely error in $\xi$ is $\pm0.5$~km~s$^{-1}$.
For Russell's data we did not have the same range of \ion{Fe}{i} lines
available. For these targets we simply adopted a mean Pleiades value of
2.3\,km\,s$^{-1}$. These microturbulence values are larger by about
1~km~s$^{-1}$ than assumed
by Boesgaard \& Friel (\cite{bf90}) or measured by King et
al. (\cite{king2000}). 
We are not concerned by this. Some of the difference can be attributed to using a
slightly higher solar microturbulence, although as King et al. used
different atmospheric models we would not necessarily expect
agreement. A consistency check on the assumed values of $T_{\rm eff}$
and $\xi$ is offered by our derived [Fe/H] values. We find a mean of
[Fe/H]\,$=-0.06\pm0.03$, in reasonable agreement with Boesgaard \&
Friel and King et al.

\begin{table}
\caption[Wavelengths and $gf$ factors for Fe {\sc i} lines used in microturbulence determination]{Wavelengths and $gf$ factors for Fe {\sc i} lines used in microturbulence determination (\ddag~and~\dag~indicate lines which are blended)}
\begin{center}
\begin{tabular}{llll}
\hline
Wavelength    & Calculated  &             & Excitation \\
({\AA})       & $gf$ factor & $gf$ factor & Potential \\
\hline
6024.049      & $-$0.250 & $-$0.120 & 4.549 \\
6027.050      & $-$1.300 & $-$1.210 & 4.076 \\
6055.992      & $-$0.550 & $-$0.460 & 4.733 \\
6062.846      & $-$4.130 & $-$4.140 & 2.176 \\
6065.805      & $-$4.168 & $-$4.168 & 3.301 \\
6078.491      & $-$0.729 & $-$0.424 & 4.795 \\
6078.999      & $-$1.200 & $-$1.120 & 4.652 \\
6136.615      & $-$1.619 & $-$1.400 & 2.453 \\
6136.993      & $-$3.202 & $-$2.950 & 2.198 \\ 
6137.270\ddag & $-$2.252 & $-$1.944 & 4.580 \\
6137.497\ddag & $-$2.601 & $-$3.263 & 3.332 \\
6137.694\ddag & $-$1.788 & $-$1.403 & 2.588 \\
6147.829      & $-$1.600 & $-$1.700 & 4.076 \\
6705.101      & $-$1.248 & $-$1.433 & 4.607 \\
6710.316      & $-$4.969 & $-$5.296 & 1.485 \\
6713.046\dag  & $-$1.653 & $-$1.889 & 4.607  \\
6713.195\dag  & $-$2.524 & $-$2.747 & 4.143 \\
6725.353      & $-$2.300 & $-$2.506 & 4.103  \\
6732.070      & $-$2.450 & $-$2.517 & 4.584 \\
6733.151      & $-$1.600 & $-$1.758 & 4.638 \\
\hline
\end{tabular}
\end{center}
\label{table-felines}
\end{table}

\begin{table}
\caption{Line list for the region around 6708.}
\begin{tabular}{llll}
\hline
Element & Wavelength & Excitation    & $\log gf$ \\ 
        & ({\AA})    & Potential (eV)&           \\
\hline
Fe 2 & 6706.880 & 5.956 & $-$9.840 \\
Si 1 & 6706.980 & 5.954 & $-$2.718 \\
Fe 1 & 6707.432 & 4.608 & $-$2.357 \\
Sm 2 & 6707.450 & 0.920 & $-$2.200 \\
 V 1 & 6707.518 & 2.743 & $-$0.165 \\
Cr 1 & 6707.644 & 4.207 & $-$2.667 \\
\bf{Li 1}$\star$ & 6707.754 & 0.000 & $-$0.431 \\
\bf{Li 1}$\star$ & 6707.766 & 0.000 & $-$0.209 \\
\bf{Li 1}$\star$ & 6707.904 & 0.000 & $-$0.773 \\
\bf{Li 1}$\star$ & 6707.917 & 0.000 & $-$0.510 \\
Fe 1 & 6708.609 & 5.446 & $-$2.298 \\
Ti 1 & 6708.755 & 3.921 &  0.094 \\
Fe 2 & 6708.885 & 10.909 & 1.458 \\
\hline
\multicolumn{4}{l}{$\star$ indicates use of ABO/tuned values for the
Van der Waals}\\ 
\multicolumn{4}{l}{broadening coefficients -- see Sect.\ref{sec-analysis}}\\
\end{tabular}
\label{table-7lines}
\end{table}

\begin{table}
\caption{Line list for the region around 6104.}
\begin{tabular}{llll}
\hline
Element & Wavelength & Excitation    & $\log gf$ \\ 
        & ({\AA})    & Potential (eV)&           \\
\hline
Co 1 & 6100.769 & 4.504 & $-$1.668 \\
Si 1 & 6102.136 & 5.984 & $-$2.120 \\
Fe 1 & 6102.159 & 4.608 & $-$2.692 \\
Fe 1 & 6102.173 & 4.835 & $-$0.454 \\
Si 1 & 6102.408 & 5.984 & $-$2.800 \\
Fe 1$\star$ & 6102.606 & 4.584 & $-$2.501 \\
 V 1 & 6102.706 & 3.245 & $-$0.751 \\
Ca 1$\star$ & 6102.723 & 1.879 & $-$0.950 \\
Ti 1 & 6102.821 & 1.873 & $-$2.315 \\
Fe 1$\star$ & 6103.186 & 4.835 & $-$0.721 \\
Fe 1$\star$ & 6103.294 & 4.733 & $-$1.325 \\
Fe 2$\star$ & 6103.496 & 6.217 & $-$2.224 \\
{\bf Li 1}$\star$ & 6103.538 & 1.848 & 0.101 \\ 
{\bf Li 1}$\star$ & 6103.649 & 1.848 & 0.361 \\
{\bf Li 1}$\star$ & 6103.664 & 1.848 & $-$0.599 \\
Si 1 & 6104.019 & 5.954 & $-$3.030 \\
Sm 2 & 6104.781 & 1.798 &  0.031 \\
\hline
\multicolumn{4}{l}{$\star$ indicates use of ABO/tuned values for the
Van der Waals}\\ 
\multicolumn{4}{l}{broadening coefficients -- see Sect.~\ref{sec-analysis}}\\
\end{tabular}
\label{table-4lines}
\end{table}

\subsection{Fitting the lines}
Once all the required model parameters were determined, lithium
abundances were obtained by fitting a synthesized spectrum to the data
using {\sc uclsyn} (Smith \cite{smith92}; Smalley et
al. \cite{smalley01}). In the case of 6708 the synthesized region was
from 6706.5{\AA} to 6709.5{\AA}. The only significant line (apart from
\ion{Li}{i}) was the (partially-resolved) \ion{Fe}{i} 6707.43\AA\
line. The 6708\AA\ line was treated as a quadruplet (see
Table~\ref{table-7lines}).  The 6104 region was modelled between
6100.5{\AA} and 6105.5{\AA}. Here, the nearby \ion{Ca}{i}, \ion{Fe}{i}
and \ion{Fe}{ii} lines were all important and the 6104\AA\ line was
synthesized as a triplet (see Table~\ref{table-4lines}). Abundances of
Fe and Ca were fixed at $solar-$0.03, or A(Fe) = 7.51 and A(Ca) = 6.33
(assuming a solar [Ca/Fe] ratio). This iron abundance was determined
spectroscopically by Boesgaard \& Friel
(\cite{bf90}). We use this value rather than our own
derived value for each star, because we believe that some of these stars
are considerably affected by large surface inhomogeneities such as star spots
-- see Sect.~\ref{sec-atms}. This might introduce errors into our [Fe/H]
values that would not have been present in the hotter F stars
considered by Boesgaard \& Friel. It might also of course introduce errors in our Li
abundances, which we will subsequently consider at some length.

The lithium
abundance was determined by varying it until a minimum value of
chi-squared was obtained; 1-$\sigma$ errors in abundance were found by
searching for abundance values that increased this chi-squared value by
1.  The EW for the Li lines were then determined from the synthetic
spectra, though we stress the abundance is determined by the synthetic
fit, not via a curve of growth. The results are given in
Table~\ref{table-results}.

The 6104\AA\ line is difficult to measure as it is very weak and lies
in the red-ward wings of strong \ion{Ca}{i} and \ion{Fe}{i} lines, and
next to a weak \ion{Fe}{ii} feature, so high resolution spectra are
required to facilitate detection. In four cases 6104 can be clearly
seen with a significant EW (Hz~2284, Hz~129, Hz~263 and Pels~19).  In
four cases the line is marginally detected. That is, there is a
reduction in $\chi^2$ when 6104 is added to the synthesis, but not at a
level sufficient to claim a secure detection.  For these objects
(Hz~746, Pels~143, Hz~1095, Pels~44) we quote the best fit abundance,
but caution the reader that the two sigma errors in abundance are
consistent with no Li at all.  In cases where chi-squared was not
improved by adding Li (Hz~522, Hz~2311 and Hz~2106), we instead
estimated upper limits to the Li abundance by producing an increase in
$\chi^2$ consistent with a 99\%-confidence interval.

\section{Results}

\begin{figure}
\resizebox{\hsize}{!}{\includegraphics{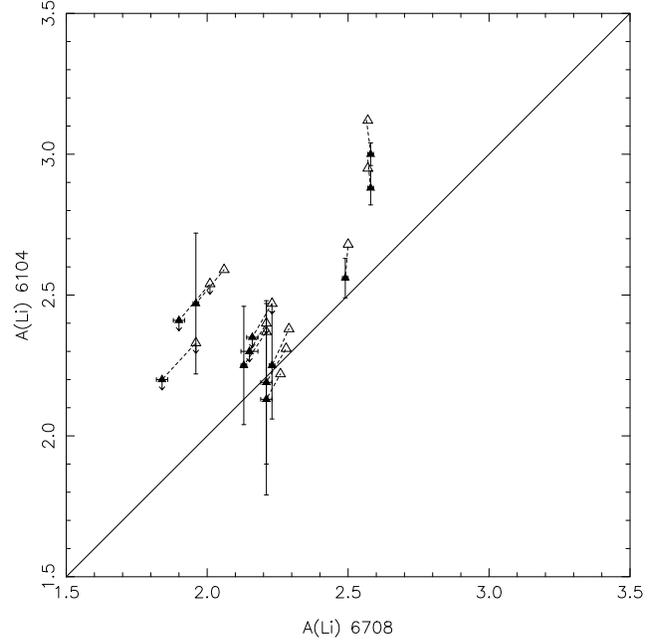}}
\caption{LTE-derived (filled triangles) and NLTE-corrected (open triangles) A(Li)$_{6104}$ {\em versus} A(Li)$_{6708}$ for sample
stars. Points for each object are connected by dotted lines.}
\label{fig-4v7plot}
\end{figure}

The Li abundances obtained from our syntheses are LTE values.  The
6104\AA\ and 6708\AA\ lines are affected by small NLTE corrections
which we apply using the the code of Carlsson et
al. (\cite{carlsson94}). Both LTE-derived and NLTE-corrected abundance
are presented, together with synthetic EWs, in Table~\ref{table-results}.
LTE-derived and NLTE-corrected abundances are plotted in
Fig.~\ref{fig-4v7plot}, with lines joining the LTE and NLTE abundances for
each object. Note that the uncertainties quoted are statistical,
relating to the fitting process, and do not include systematic errors,
which we discuss below.

\begin{table*}
\caption{Synthetic Li EWs, EW of the core of H$\alpha$, LTE-derived Li abundances, and NLTE-corrected Li abundances for sample spectra.}
\begin{tabular}{llrrrrrrrr}
\hline
Object & $T_{\mathrm{eff}}$ & H$\alpha$ EW & EW$_{6708}$ & A(Li)$_{6708}$ & A(Li)$_{6708}$ & EW$_{6104}$ (m{\AA}) & A(Li)$_{6104}$ & A(Li)$_{6104}$ & $\Delta$A(Li)$^{\star}$ \\ 
& & (\AA) & (m\AA)  & LTE & NLTE  & (m\AA) & LTE & NLTE &  \\
\hline
Hz 2284  & 5449 & $0.66\pm0.02$ & 138.9$\pm$2.1 & 2.58$\pm$0.01 & 2.57 &13.6$\pm$1.6  & 2.88$\pm$0.06 & 2.98       & +0.41  \\
 Hz 746  & 5383 & $0.60\pm0.02$ &  86.2$\pm$2.2 & 2.21$\pm$0.02 & 2.25 & 2.8$\pm$1.7  & 2.13$\pm$0.34 & 2.22       & $-$0.03  \\
Hz 2106  & 5245 & $0.47\pm0.02$ & 103.9$\pm$2.5 & 2.16$\pm$0.02 & 2.23 &$\leq$5.8     & $\leq$2.35    & $\leq$2.47 & $\leq$0.24 \\
 Hz 129  & 5228 & $0.46\pm0.02$ & 164.5$\pm$0.9 & 2.49$\pm$0.01 & 2.50 & 8.2$\pm$1.5  & 2.56$\pm$0.07 & 2.68       & +0.18  \\
 Hz 263  & 5213 & $0.40\pm0.02$ & 185.9$\pm$1.9 & 2.58$\pm$0.01 & 2.57 &25.0$\pm$2.5  & 3.00$\pm$0.04 & 3.12       & +0.55  \\
 Hz 522  & 5110 & $0.61\pm0.02$ & 78.5$\pm$2.1 & 1.84$\pm$0.02 & 1.96 & $\leq$5.0  & $\leq$2.20      & $\leq$2.33 & $\leq$0.37 \\
\hline
Hz 2311  & 5352 & $0.65\pm0.04$ &  83.5$\pm$3.6 & 2.15$\pm$0.03 & 2.21 &$\leq$4.3     & $\leq$2.30   & $\leq$2.40 & $\leq$0.19 \\
Pels 143 & 5177 & $0.66\pm0.03$ & 126.0$\pm$2.7 & 2.21$\pm$0.02 & 2.28 &4.4$\pm$2.3  & 2.19$\pm$0.29 & 2.31       & +0.03  \\
Pels 19  & 5172 & $0.59\pm0.02$ &  84.7$\pm$1.5 & 1.96$\pm$0.01 & 2.06 &8.4$\pm$4.1  & 2.47$\pm$0.25 & 2.59       & +0.53  \\
Hz 1095  & 5160 & $0.61\pm0.02$ & 131.6$\pm$1.8 & 2.23$\pm$0.01 & 2.29 &5.2$\pm$2.1  & 2.25$\pm$0.19 & 2.38       & +0.09 \\
Pels 44  & 5167 & $0.66\pm0.02$ & 113.6$\pm$1.7 & 2.13$\pm$0.01 & 2.21 &5.2$\pm$2.0  & 2.25$\pm$0.21 & 2.38       & +0.17 \\
Hz 522   & 5110 & $0.65\pm0.02$ & 86.7$\pm$2.3 & 1.90$\pm$0.02 & 2.01 &$\leq$8.1  & $\leq$2.41      & $\leq$2.54 & $\leq$0.53  \\
\hline
\multicolumn{9}{l}{$^{\star}$ this is A(Li)$_{6104}$ - A(Li)$_{6708}$ for NLTE abundances}
\end{tabular}
\label{table-results}
\end{table*}

There is an apparent scatter in Li abundance from both lines, with a
spread of $\sim$0.7\,dex for each line, although this could potentially
be larger for 6104, due to the presence of upper limits. The abundances
measured from the two lines do at least appear to be correlated. However, before we
can accept this scatter as real, we need to consider possible sources
of error in the synthetic abundance determinations.

\subsection{Atomic parameter uncertainties}
\label{sec-fits}

When fitting the Li lines, the main sources of error (aside from the
statistical errors) which are likely
to arise from the measurement process are the atomic parameters:
damping and oscillator strengths.

Stark and Van der Waals damping effects arise from perturbations due to
charged particles (Stark) and neutral hydrogen (Van der Waals) and
shape the wings of the lines. Stark broadening is weaker than Van der
Waals effects in the outer layers of the atmosphere, becoming important
only deep within the star. Total removal of the Stark broadening
affects derived Li abundances by only $\sim$0.001\,dex, whilst neglect
of Van der Waals effects in the Li lines can lead to abundance
variations of 0.022\,dex for 6708, or 0.001\,dex for 6104. The broadening
of lines near Li, especially around 6104, can also affect the derived
abundances, since the wings of the strong \ion{Ca}{i} and \ion{Fe}{i}
lines contribute to absorption at the Li wavelength.  In our syntheses we have used
the best available (ABO) Van der Waals co-efficients (Barklem et
al. \cite{barklem00}) for the \ion{Li}{i}, and \ion{Ca}{i} lines and tuned these
coefficients for the \ion{Fe}{i} lines to get a good fit to the solar atlas.
Using the older Kurucz \& Bell (1995) coefficients would
result in approximately 0.05-0.25\,dex higher Li abundances from the
6104\AA\ line, where the smaller increase is appropriate for our
targets with the largest Li abundance. The increase is mainly due to
the broad wings of the very strong \ion{Ca}{i} line being interpreted
as extra absorption due to Li.

The other source of atomic uncertainties is the oscillator
strengths. These are difficult to measure reliably from astrophysical
spectra. However, since we are not interested in the Ca and Fe lines,
other than to ensure we get the best fits to the Li features, we have
fitted the $gf$s for these other elements using a solar atlas and
observed spectra (see Sect.~\ref{sec-analysis}). We did not fit the $gf$s
for either of the Li lines, since the Li abundance in the Sun is low
enough that the lines are barely visible, even in the high-resolution
solar atlas. The values we have used for Li are those given in the
Vienna Atomic Line Database (VALD -- Piskunov et al \cite{piskunov95})
which is the best source for such data at the present time. The Li $gf$
values given in VALD are also the same as those presented by Lindgard
\& Nielsen (\cite{lindgard77}). It is possible that the 6104 $gf$
factors could have been wrongly estimated, which would affect the
abundances derived from the line: uncertainties in the $gf$ values are
quoted at $<$3\% for 6708 and $<$10\% for 6104
(corresponding to abundance uncertainties of about 0.01 and 0.04
dex respectively) by Weiss (\cite{weiss63}). We also determined a $gf$
factor for the \ion{Fe}{ii} line adjacent to 6104, with the new value
being 0.053\,dex smaller than the value quoted in the original line list. The
change in $gf$ has a negligible impact on the 6104-derived Li
abundances, although the inclusion of the \ion{Fe}{ii} line in the
synthesis, {\em is} significant (see below).

\subsection{Atmospheric uncertainties}
\label{sec-atms}

Other than those uncertainties due to fitting the lines, there are
likely to be systematic uncertainties in the determined abundances due
to the atmospheric model parameters used. These might change the comparison of
abundances derived from 6104 and 6708 or indeed inject scatter into the
derived abundances from both lines. To gain an understanding of the
effects that changes of $T_{\rm eff}$, $\log g$, $\xi$ and the
metallicity of the atmosphere, we attempted to fit our spectra after
perturbing each of these parameters in turn. The new Li abundances were
then compared with the original fits to see what size abundance errors
could result from plausible changes in the atmospheric parameters.
Note that this procedure is more complex than simply re-generating
curves of growth for each atmospheric model (which is often done in the
literature). We find that, especially in the case of 6104, the
deduced Li abundances vary by more than would be expected from a simple
curve of growth, because of the interplay with the wings of the strong
adjacent lines and the presence of the nearby \ion{Fe}{ii} line.

Example results from these tests are listed in
Table~\ref{table-errors}, for two stars (Hz~263 and Pels~44) that are
representative of the highest and lowest Li abundances found in our
dataset.  Of the atmospheric parameters, changes in $T_{\mathrm{eff}}$
have the most significant effect on abundance. The photometry errors
were small, $\sim$0.02 (Soderblom et al. \cite{s93a}), propagating to a
temperature difference of about 50\,K at 5300\,K.  A change in model
temperature will affect abundances from both lines to almost the same
extent and systematically in the same direction -- about 0.12\,dex per
100\,K change in $T_{\rm eff}$. The change in the 6104-derived
abundance is about twice what one would expect from a simple
consideration of the 6104 curve of growth, but appears more or less
independent of the strength of the Li feature.  The effects of changing
$\xi$, $\log g$ and [Fe/H] by plausible amounts are much smaller,
except in the case of very weak 6104-derived abundances.  Here we find
that the blending wings of the strong adjacent lines {\em are}
sensitive enough to these parameters that they lead to additional Li
abundance errors.  It is worth noting here that assuming
systematically lower $\xi$ values for the Sun and Pleiades stars would
make no systematic difference to the 6104-derived abundances, because
we would simply have increased the damping widths of the nearby lines
to compensate and match the solar spectrum. There would be a very small
systematic increase in the 6708-derived abundances, but by no more than indicated in
Table~\ref{table-errors} for a small $\xi$ change.

We can conclude from these considerations that
plausible uncertainties in the atmospheric parameters {\em cannot}
explain the observed $\geq0.7$~dex scatter in the abundances measured
from both lines.

A further source of error is likely to be the assumption of a
single-temperature, plane-parallel, homogeneous atmosphere. This is
discussed at some length in Sect.~\ref{sec-spots}. Here we note that
the fits to some (but not all) of the Pleiades stars appear rather poor
in the cores of the Fe and Ca lines around 6104\AA. However, the wings
of the lines seem to be well fitted, and rather than arbitrarily change
the Ca and Fe abundances by considerably more than 0.1\,dex (for which
there is no physical motivation), we leave the abundances at the mean
Pleiades value. The lack of a good fit in the cores of some of these
lines is clear evidence that, at least for some stars, something is
lacking in our atmospheric models. This is also clearly seen in
some of the Ca 6718\AA\ lines in Fig.~\ref{fig-spec74} in the form of
central reversals or asymmetries. As this line is often used for
doppler imaging of star spots on cool stars we take these anomalous
spectral features to indicate the presence of large spotted regions 
on at least some of our stars.

\begin{table*}
\caption[Effects of different atmospheric parameters on A(Li)]{Effects
of different parameters on A(Li). Results are given for Hz~263 and Pels
44, assuming the baseline atmospheric parameters listed in Table~\ref{table-data1}.
The first row gives the unperturbed LTE abundances.
Subsequent lines give the abundance for the stated parameter
variation relative to the initial model. We do not expect the
abundances to be accurate to 3 decimal places, but these are included
to illustrate the small scale of some of the variations.}
\small
\begin{center}
\begin{tabular}{llrrrr}
\hline
          &           & \multicolumn{2}{c} {Hz~263} &
          \multicolumn{2}{c} {Pels 44} \\  
Parameter             & Varies by   & 6708{\AA} & 6104{\AA} & 6708{\AA} & 6104{\AA} \\
\hline
 ---	              & ---         & 2.584    & 3.000   & 2.128  & 2.250     \\
\hline
$T_{\mathrm{eff}}$    & $-$100\,K    & $-$0.125 & $-$0.125 & $-$0.121 & $-$0.125 \\
$T_{\mathrm{eff}}$    & $+$100\,K    &  +0.121 &  +0.094 &  +0.117 &  +0.125 \\
$\log g$              & +0.3        &  -0.004 &  0.000  & $-$0.002 & $-$0.141 \\
$[$Fe/H$]$            & +0.1        &  +0.008 &  0.000 & $-$0.003 & +0.062 \\
$\xi$                 & +0.5~km~s$^{-1}$ & $-$0.024 & $-$0.008 & $-$0.003  & -0.062 \\
\hline
\end{tabular}
\label{table-errors}
\end{center}
\normalsize
\end{table*}

\subsection{Comparison with Russell's results}

The observations of Russell's data were taken under similar conditions
to our own, except that he used an EEV CCD which has a smaller pixel
size than the SITe 1, and a different central wavelength.  
Russell's analysis differs from our own in a number of ways. The most
straightforward is in the temperature
calibrations used. He makes use of the calibration of Bessell (1979):
\begin{eqnarray}
T_{\mathrm{eff}} = 8899 - 6103(B-V)_{\mathrm{o}} + 1808(B-V)^{2}_{\mathrm{o}}
\end{eqnarray}
which was also used by Soderblom et al. (1993a). Our temperature
calibration is that of B\"ohm-Vitense (\cite{bv81}). This matches that
used by Russell at 6000\,K, however for the cooler stars considered here our temperatures
are higher by $\sim$150\,K. Following our consideration of abundance
errors from $T_{\mathrm{eff}}$ uncertainties (Sect.~\ref{sec-atms}) we
see that this difference should result in Russell obtaining smaller
6708-derived abundances (by $\sim$0.18\,dex) for a given EW. The
6104-derived abundance will also be smaller by a similar amount.
Table~\ref{table-russ-us} contains the abundances obtained both in this paper
and by Russell, and the temperatures used in each analysis.

Our NLTE-derived 6708 abundances are systematically higher than those
reported by Russell, by an average of 0.2\,dex (using our temperatures),
which suggests that the majority of the difference in these abundances
is due to the different temperature calibrations used. There are also
different measurement techniques to be considered. Russell simply
fitted a Gaussian to 6708 and we can confirm Russell's EWs when
we model these spectra in the same way. However, our synthetic EWs 
are systematically {\em smaller} than reported by
Russell, by an average of 10.7m{\AA}. This appears to be mainly because
the feature is not Gaussian. In particular, the
\ion{Fe}{i} line, blue-ward of the Li line is {\em not} adequately
deblended in the Gaussian fit. Allowing for this discrepancy in EW, our
abundances are still about 0.07\,dex higher than Russell's, which we
attribute to his use of an older version of Kurucz's Atlas atmospheres.
An exception to this is Hz~2311 (which was not included when
calculating the mean abundance and EW differences mentioned above), where our abundance
is lower than that determined by Russell, and our EW value for the star
is much lower than Russell reported. Why this should be the case is
unclear, although this object has the lowest S/N of any in our sample,
so there could be some major differences in continuum placement between
the two objects.

6104 abundances determined using our temperatures are {\em smaller}
by 0.1 to 0.3\,dex, and our synthetic 6104 EWs seem to be
smaller by 2-6 m\AA\ than those deduced by Russell. The abundance
discrepancy would be about 0.18\,dex wider if we had used Russell's
temperature scale. The exception is Pels~19, where our EW (and
abundance) are both larger than previously reported. Russell did not
publish his reduced spectrum for this object, so it is difficult to
ascertain where such a discrepancy could have arisen. 6104 is
present in the spectrum of Pels~19 (see Fig.~\ref{fig-russspec}), where no line
is clearly visible in Pels~143 (even though Russell reports a larger EW for
Pels~143).  We have also obtained only upper limits to the abundance
for Hz~2311 and Hz~522, where Russell claimed to have detected 6104.

We believe that these important differences, upon which our conclusions
(and those of Russell) hinge, can be traced to the comprehensiveness of
the fitting and synthesis techniques employed.  In the case of 6104,
Russell used two different techniques, both of which are open to
criticism: fitting all the (strong \ion{Fe}{i} and \ion{Ca}{i} in
addition to \ion{Li}{i} 6104) lines around 6104 simultaneously with
Gaussians; and fitting Gaussians to the (strong) Fe and Ca lines near
Li 6104 (but omitting the weak \ion{Fe}{ii} line), subtracting the
synthesis from the spectrum and then fitting the residual with another
Gaussian. We believe this neglect of \ion{Fe}{ii}, along with the
non-Gaussian nature of the surrounding strong, damped lines, could
easily result in overestimates of the 6104 EW and Li abundance. As the
6104 line is also a triplet, it is unlikely to be well represented by a
Gaussian, and the red-most \ion{Fe}{i} feature is actually a blend of
two lines (this can just be seen in the solar atlas of this region --
Fig.~\ref{fig-solar}).

We gauged the magnitude of these problems by running the synthesis for
a line list containing (a) the \ion{Fe}{i}, \ion{Ca}{i} and
\ion{Fe}{ii} lines and (b) the \ion{Fe}{i} and \ion{Ca}{i} lines only,
in addition to the Li triplet. The Li abundances obtained in the
absence of the \ion{Fe}{ii} line appear to show a trend: the abundance
being {\em overestimated} by an amount which increases for the objects
with smaller Li abundance (from the 6708 line). We would obtain Li
abundances that are higher by between 0.02 and 0.10\,dex as a result of
neglecting the \ion{Fe}{ii} line, which would turn our upper limits
into marginal detections.  Omitting other weak lines from the list
(while including \ion{Fe}{ii}) has a negligible effect on the abundance
and EW attributable to 6104.  The neglect of the damping wings on the
Li abundance can be judged from our change from the Kurucz \& Bell
(1995) Van-der-Waals coefficients to the larger values required to fit
the solar atlas. The lines are more closely approximated by Gaussians
in the former case and as mentioned in the last section, result in Li
abundances about 0.05-0.25\,dex higher than when the broader line wings
are used. Again, the biggest effect is for the objects with the weakest
Li.  We conclude that both the neglect of the \ion{Fe}{ii} line and the
(implicit) neglect of the broad damping wings by assuming Gaussian line
profiles conspire to produce larger 6104-determined Li abundances, such
that the lowest Li abundances would be increased the most. This would
raise the average Li abundance and reduce the apparent abundance
scatter deduced from this line. It might also lead to claims of
detection for 6104, where in reality the line was too weak to be
seen. It is worth noting however, that Russell's conclusion that the spread in 6708-derived Li abundances was larger than that from 6104 can also be explained, in part, by the much larger 6708-derived Li
abundance found for Hz 2311 by Russell.

\begin{table*}
\caption{$T_{\mathrm{eff}}$ values and NLTE-derived Li abundances from (columns 2-4) this paper and (columns 5-7) Russell's published results.}
\begin{tabular}{lcrrcrr}
\hline
 & \multicolumn{3}{c}{This paper}& \multicolumn{3}{c}{Russell(1996)} \\
Object & $T_{\mathrm{eff}}$ (K) & EW (m{\AA}) & A(Li) & $T_{\mathrm{eff}}$ (K) & EW (m{\AA}) & A(Li) \\
\hline
\multicolumn{7}{c}{6708}\\
\hline
Hz 2311  & 5352 &  83.5$\pm$3.6 & 2.21$\pm$0.03 & 5240 & 141$\pm$5.6 & 2.37 \\
Pels 143 & 5177 & 126.0$\pm$2.7 & 2.28$\pm$0.02 & 5020 & 139$\pm$4.2 & 2.09 \\
Pels 19  & 5172 &  84.7$\pm$1.5 & 2.06$\pm$0.01 & 5020 &  95$\pm$2.8 & 1.85 \\
Hz 1095  & 5160 & 131.6$\pm$1.8 & 2.29$\pm$0.01 & 5000 & 139$\pm$3.8 & 2.08 \\
Pels 44  & 5167 & 113.6$\pm$1.7 & 2.21$\pm$0.01 & 4990 & 123$\pm$3.8 & 2.01 \\
Hz 522   & 5110 &  86.8$\pm$2.3 & 2.01$\pm$0.02 & 4940 & 100$\pm$4.3 & 1.79 \\
\hline
\multicolumn{7}{c}{6104}\\
\hline
Hz 2311  & 5352 & $\leq$4.3     & $\leq$2.40    & 5240 &  7.7 & 2.53 \\
Pels 143 & 5177 & 4.4$\pm$2.3   & 2.31$\pm$0.29 & 5020 &  7.2 & 2.37 \\
Pels 19  & 5172 & 8.4$\pm$4.1   & 2.59$\pm$0.25 & 5020 &  6.2 & 2.31 \\
Hz 1095  & 5160 & 5.2$\pm$2.1   & 2.38$\pm$0.19 & 5000 & 10.2 & 2.50 \\
Pels 44  & 5167 & 5.2$\pm$2.0   & 2.38$\pm$0.21 & 4990 & 11.2 & 2.55 \\
Hz 522   & 5110 & $\leq$8.1     & $\leq$2.41    & 4940 & 11.0 & 2.50 \\
\hline
\end{tabular}
\label{table-russ-us}
\end{table*}

\subsection{6104 {\em versus} 6708}

\begin{figure}
\resizebox{\hsize}{!}{\includegraphics{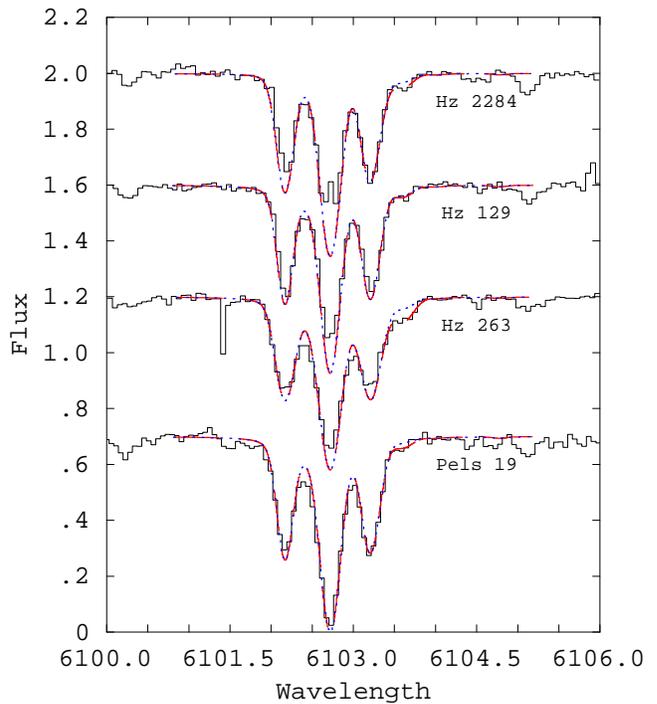}}
\caption{Spectra of objects with firm 6104 detections (solid lines) in
the region of the 6104{\AA} \ion{Li}{i} line, showing LTE syntheses for
abundances determined from 6104 (dashed) and 6708 (dotted) lines.}
\label{fig-4-7}
\end{figure}

In addition to scatter in Li abundances from both 6708 and 6104, we
find that some 6104-derived NLTE-corrected abundances are higher than
the corresponding 6708 values. The four objects where this is clearly
the case are: Hz~2284 (A(Li)$_{6104-6708}$ = 0.41$\pm$0.06\,dex), Hz~129
(0.18$\pm$0.07\,dex), Hz~263 (0.55$\pm$0.04\,dex) and Pels~19
(0.53$\pm$0.25\,dex). These stars are in fact the only ones where
the 6104\AA\ line is clearly detected.
To emphasize these discrepancies, in Fig.~\ref{fig-4-7} we show LTE
syntheses for the 6104 region using the LTE Li abundance suggested from
the 6708 line and the LTE Li abundance fitted to the observed 6104
region.  Temperature errors are probably not the culprit here because a
100\,K temperature increase causes A(Li)$_{6104}$ to increase by around
0.12\,dex, but affects the 6708-derived abundances in a similar way. As
discussed in Sects.~\ref{sec-fits} and~\ref{sec-atms}, the differential
abundance errors due to uncertainties in $\xi$, $\log g$ and the $gf$
values are probably quite small in these stars, though not negligible.
This abundance discrepancy {\em might} be present in the other stars, once the large
errors in the 6104-derived abundance are taken into account, but
equally, the data are also consistent with good agreement between
abundances determined from 6104 and 6708 for these stars.

\begin{figure}
\resizebox{\hsize}{!}{\includegraphics{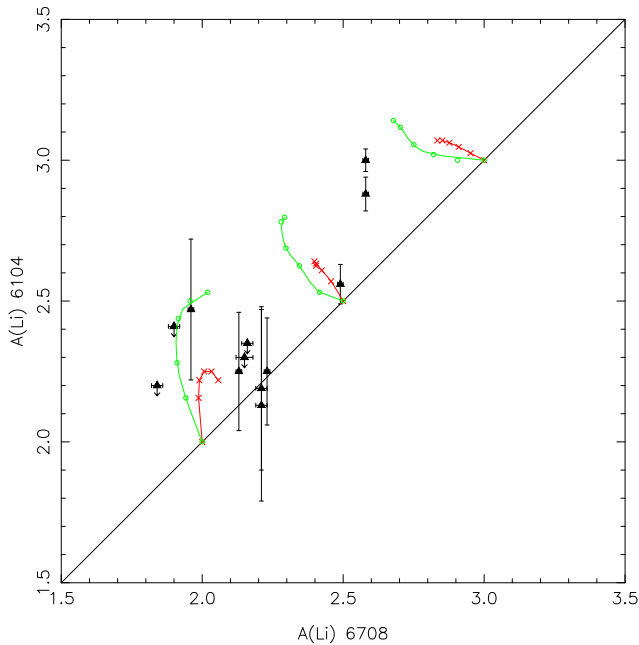}}
\caption{LTE A(Li)$_{6104}$ {\em versus} A(Li)$_{6708}$ for sample stars. The curves represent the effect star spots would have on the abundances for a star with $T_{\mathrm{eff}}$\,=\,5250\,K. Crosses represent the case where the spotted area is 1000\,K cooler than the stellar surface, while circles show regions where the temperature difference is 1500\,K. The symbols mark spotted areas of 0\%, 10\%, 20\%, 30\%, 40\% and 50\%. See also Sect.\ref{sec-spots}.}
\label{fig-lte2nlte}
\end{figure}

The difference between the 6708 and 6104 abundances has been
exacerbated by the application of NLTE corrections, as can be seen in
Fig.~\ref{fig-lte2nlte}. At the effective temperatures of our targets, and
for low Li abundances, the NLTE corrections affect both lines roughly
equally, increasing the determined abundances by 0.1\,dex. At larger
abundances, the NLTE corrections to the 6708 abundance become small and
then reverse sign, whereas the 0.1\,dex correction to the 6104 abundance
remains roughly constant (see Carlsson et al. \cite{carlsson94}).

We conclude, on the basis of the models that are available to us, that
there is broad agreement and at least a correlation between the 6104
and 6708 abundances. However for some, but not necessarily all, of the
stars we have investigated, the 6104-determined Li abundance gives
higher values (by 0.2-0.5\,dex) than that from 6708. 

\section{Discussion}
\label{sec-spots}
The main aim of this study was to investigate whether the Li abundances
determined from the subordinate line at 6104{\AA} agreed with those
derived from the 6708{\AA} resonance line. Such an agreement would lend
confidence to the atmospheric approximations and NLTE corrections used
in Li studies, and might imply that the spread in abundance previously
reported from the 6708 line alone was indicative of a physical
mechanism which could produce star-to-star differences in Li abundance
during the approach to the ZAMS.

In our analysis, we find a scatter of $\sim$0.7\,dex from {\em both}
6708 and 6104 (although the 6104 spread could be larger still, due to
upper-limit estimates for some of the objects). The scatter is not
explained by the uncertainties considered for either line. We find a
broad correlation between the abundances from the two lines although
for some stars there is an indication that the 6104-determined
abundance is still 0.2-0.5\,dex higher. The reason our results do
not entirely agree with those of Russell (1996), who found that the
6104-derived abundances were higher and showed less scatter than those
from 6708, can be traced to: (a) The large 6708-derived abundance for
Hz 2311 found by Russell, which we have not been able to confirm. This
increases the scatter in Russell's 6708-derived abundances; and (b)
Russell's neglect of the strong, non-Gaussian damping wings of lines
that blend with 6104 and the neglect of a weak \ion{Fe}{ii} line, both
of which might have been interpreted as Li absorption.

Although the abundance scatter derived from both lines supports the
claim that the abundance spread is real, the disagreement in abundance
for some cases does still illustrate a deficiency in our understanding
of the stellar atmospheres.  Due to their ease of computation,
one-dimensional atmospheres have been used for most abundance
determinations until recently. However, they do not allow for the
existence of inhomogeneities such as star spots on stellar surfaces,
which are certainly known to exist in young, magnetically active,
late-type stars and particularly in the Pleiades G and K stars
(e.g. Krishnamurthi et al. \cite{krish98}).  Doppler imaging study of
the Pleiades G-type star (T$_{\mathrm{eff}}$~=~5845\,K) \object{Hz~314}
(Rice \& Strassmeier \cite{rice01}) showed that the star had cool spots
at or near the pole, and within the equatorial regions. The
temperatures of the spots varied from 4400\,K to 5400\,K, with the
average being $\sim$4700\,K.  Spot areas have been measured
spectroscopically for several highly-active G and K stars by O'Neal et
al. (\cite{oneal98}), who find spot coverage of 3-56\%, and temperature
differences between spotted and `quiet' (unspotted) regions ranging
from 750\,K to 1900\,K.
 
The presence of spots could alter our conclusions in two ways: they
might reduce the spread in Li abundances significantly in a group of stars with
differing spot coverage (this has previously been explored by Soderblom
et al. \cite{s93a} and Barrado y Navasc\'ues et al. \cite{barrado01},
who conclude that the effect is not large enough). However, 6104 and
6708 might be affected differently due to their differing curve of
growth temperature sensitivities, so differing spot coverage might well
explain why some stars have agreement between 6708- and 6104-derived
abundances while others reveal a higher abundance from 6104.

In order to probe
what effect star spots might be expected to have on the abundances
determined in this paper, we used a two-component, one-dimensional atmospheric
simulation (TCODAS) technique included in the {\sc uclsyn} code.

In the TCODAS models we balanced the spot and unspotted (quiet) star temperatures and areas, so that the luminosity-weighted average
$T_{\rm eff}$ would remain the same, using
\begin{eqnarray}
aT_{\mathrm{eff}}^{4} = a_{\mathrm{cool}}T_{\mathrm{cool}}^{4} +
a_{\mathrm{star}}T_{\mathrm{star}}^{4}
\label{eqn-tcool-teff}
\end{eqnarray}
where $a_{\mathrm{cool}}$ and $T_{\mathrm{cool}}$ are the cool-region area and temperature, and $a_{\mathrm{star}}$ and $T_{\mathrm{star}}$ are the area and temperature for the rest of the star. The areas are normalised to unity, such that:
\begin{eqnarray}
a_{\mathrm{star}} + a_{\mathrm{cool}} = 1
\end{eqnarray}
$\Delta T$ (= $T_{\mathrm{star}} - T_{\mathrm{cool}}$) and
$a_{\mathrm{cool}}$ were fixed at a number of discrete points, then
Eq.~\ref{eqn-tcool-teff} was solved for the appropriate $T_{\mathrm{eff}}$. We assume that the cool regions are uniformly distributed across the star to avoid difficulties in modelling line-profile asymmetries. 

The relative-light ($LR$) contributions from the two components were flux weighted using
\begin{eqnarray}
LR_{\mathrm{cool}} = \frac{a_{\mathrm{cool}} \times
T_{\mathrm{cool}}^4}{T_{\mathrm{eff}}^4}
\end{eqnarray}
and
\begin{eqnarray}
LR_{\mathrm{star}} = \frac{a_{\mathrm{star}} \times
T_{\mathrm{star}}^4}{T_{\mathrm{eff}}^4}
\end{eqnarray}

We generated synthetic (LTE) spectra (at 6104 and 6708) using two
component models with a luminosity-weighted average $T_{\mathrm{eff}}$
of 5250\,K, spot coverage of between 10 and 50 percent of the visible
hemisphere and a temperature difference of either 1000\,K or 1500\,K
between spotted and unspotted components. Finally we fitted our
two-component syntheses with one component models at both 6104 and
6708. The resulting abundances are plotted on
Fig.~\ref{fig-lte2nlte}. These suggest that the abundances for Hz~263,
Hz~2284 and Pels~19 (and Hz~129) could potentially be explained by the
presence of star spots on these objects with coverage fractions of
10-50 percent and $\Delta T$ of 1000-1500\,K.  Note that the results
obtained from TCODAS are LTE abundances, since we do not have access to
NLTE syntheses, and application of 1-D NLTE corrections to the results
would most likely not be valid.

Is there any evidence to support the notion that that those stars
with significant discrepancies between their 6708- and 6104-derived
abundances are the most magnetically active and spotted?
We have no direct information
on this and we must recognize that spot coverage might change with
time, so unless observations of magnetic-activity indicators are
co-temporal, they should be treated cautiously. There are
indications of spot activity in the line profiles of Hz~263, Hz~2284 and Hz~129 (see Figs.~\ref{fig-spec74}, \ref{fig-russspec} and
\ref{fig-4-7}. Several line cores are filled in,
especially the \ion{Ca}{i} lines at 6102\AA\ and 6718\AA. It is just
such signatures which are used to make doppler maps of starspot
distributions on more rapidly-rotating stars.
However, we see little evidence of this in Pels~19, and
similar signs of spot activity are present in Hz~2311 and Hz~522
(our observation rather than Russell's), where the 6708- and
6104-derived Li abundances {\em might} be consistent (although the
constraints are not strong and discrepancies of 0.2-0.3 dex cannot be
ruled out).

A more quantitative magnetic-activity indicator could be the X-ray
luminosity. Unfortunately, measurements are not available for all our
stars and X-ray luminosity could easily vary by factors of 2-3 over
time. A better diagnostic is the strength of H$\alpha$ measured from
the same spectra as the Li abundances. Soderblom et al. (\cite{s93b}) have investigated H$\alpha$ emission in the Pleiades. They found that late-G and early-K stars exhibit H$\alpha$ absorption with a chromospherically-filled core. The amount of filling correlates with the rotation rate and other activity indicators. The strength of the underlying photospheric
absorption is a function of $T_{\rm eff}$. If, however, we make the
reasonable assumption that this intrinsic absorption is uniform across
the narrow $T_{\rm eff}$ range considered here, then we can use the EW
of the H$\alpha$ core (1\AA\ either side of the line centre) to rank
our targets in order of magnetic activity. These H$\alpha$ core EWs are
listed in Table~\ref{table-results}. All of the targets exhibit
H$\alpha$ absorption, so it is the stars with the smallest EWs that are
the most active -- namely, Hz~263, Hz~129 and Hz~2106. Two of these
have discrepant 6708- and 6104-derived abundances, whilst the third might
not have. However, Hz~2284 is ranked as one of the least active stars
by this method, yet it does show an abundance discrepancy from the two
lines, along with signs of spot activity in its Ca\,{\sc i} line profiles.

In summary the supporting evidence for a direct relationship between
star spots and a discrepancy in Li abundance measured from the resonance
and subordinate lines is ambiguous at present. The most active stars do
seem to exhibit a discrepancy but there are counter examples as well.
It should be made clear that the `spot' areas and temperature
differences are not rigorously determined, and so should be treated
only as fitting factors for use with the models, and not as actual
areas or temperatures. We also remind the reader that changes in EW
(and therefore changes in derived abundance) due to spot coverage (and
spot coverage itself) vary over time. Jeffries et al. (\cite{j94})
showed that 6708 was modulated by star spots during a rotation cycle in
\object{BD+22 4409}, a young, active star. The line strength was
observed to vary by 10$\pm$3\% (equivalent to an abundance variation of
$\sim$0.15\,dex within a one-component model) over the 10-hour rotation
period. Over longer time-scales, Jeffries (1999) reported no
variability larger than 0.1\,dex on one-year timescales in a group of
Pleiades K stars, but considered that there might be variations of
0.2-0.3\,dex over 10 years. Finally, we note that there are differences
in the two observations of Hz~522 used in this work: the lines in the
spectra obtained by Russell are stronger.  This is consistent with an
increase in spot coverage between the two sets of observations,
resulting in an {\em apparent} abundance decrease of 0.06$\pm$0.02\,dex
between 1993 and 1998, assuming a one-component atmospheric model. If
this were typical of the magnitude of this effect, it is clear that
only a small fraction of the Li abundance spread could be accounted for
in this way. Star spots could however, account for the majority of the
discrepancies between abundances derived from 6104 and 6708 which arise
due to the use of single-component models.

The fact that we can obtain agreement between the 6708- and
6104-derived abundances with relatively minor and plausible changes to
the atmospheric structure (namely the introduction of star spots)
without altering the size of the scatter observed in both the 6708 and
6104 abundances might support the idea that this abundance scatter is
real. However, it is still possible that this agreement is coincidental
and major revisions to our understanding of these young stellar
atmospheres will be required. A cautionary note should be that scatter
has also been observed in the potassium abundances of Pleiades stars
from the \ion{K}{i}~7699\AA\ line, which is formed under similar
conditions to 6708 (Soderblom et al. 1993; Stuik et al. 1997; Jeffries
1999; King et al. 2000). As there are no plausible physical mechanisms
to produce potassium depletion in these objects, this result still
points to a deficiency in the atmospheric modelling. King et al. point
to chromospheric activity altering the ionisation balance in the line
formation region as being culpable. 6104 is formed a little deeper in
the atmosphere than 6708, although there is plenty of overlap. It is
beyond the scope of this paper to investigate whether plausible changes
in the ionisation balance might also be able bring the 6104- and
6708-determined abundances into agreement, whilst at the same time
significantly reducing the apparent scatter in Li and/or K abundances,
though such an investigation is now called for.

\section{Summary}

We have obtained high resolution spectroscopy of 
a sample of 11 Pleiades late-G and early-K stars,
covering a relatively narrow range of effective temperatures, of which
six were observed by Russell (1996). These spectra were consistently reduced
and analysed. One object, Hz~522, was included in both sets of
observations.

\begin{itemize}

\item We measured the \ion{Li}{i}~6708{\AA} resonance line in all the
objects and determined Li abundances by fitting atmospheric syntheses. 
We obtained NLTE-Li abundances which were systematically
higher than those reported by Russell, by an average of 0.2\,dex. Most
of this difference could be explained by the different temperature
calibrations used in the two analyses. When the same temperatures
reported by Russell were used, the difference between reported
abundances for the lines was reduced to about 0.05\,dex.

\item We have also been able to measure Li abundances using the weak,
subordinate \ion{Li}{i}~6104\AA\ line in eight of our sample objects and
have found upper limits for the remainder. The LTE abundances (or upper
limits) derived from the 6708\AA\ and 6104\AA\ lines are reasonably
well correlated, and agree within their errors, for most of our
targets. However, for several objects the abundance derived from
6104\AA\ is significantly higher than from 6708\AA. NLTE corrections
act to increase this discrepancy. For the sub-sample observed by
Russell, we find that his multiple-Gaussian fitting technique has
overestimated the strength of the 6104 line when compared with our
spectral synthesis. This can be attributed to the the presence of a
weak \ion{Fe}{ii} line close to the 6104\AA\ line, and the broad
damping wings of the nearby \ion{Ca}{i} and \ion{Fe}{i} lines. Neglect
of these leads to Li abundance overestimates of 0.05-0.35\,dex, such
that the overestimates are larger for those objects with weaker Li
lines. It can also lead to the detection of an Li feature where none is
present. This could account for the reduced scatter in Li abundances
from the 6104\AA\ line (compared with the 6708\AA\ line) reported by
Russell.

\item A Li abundance scatter was observed from both lines: of order
$\sim$0.7\,dex for 6708, and at least 0.7\,dex for 6104, although this could
conceivably be larger due to the presence of upper limits. We
found no evidence of a reduced scatter in the abundances determined from
6104, contrary to the conclusions of Russell (see above).  The
implications of this spread are either that there is real star-to-star
scatter present, or that the agreement between the lines is
coincidental and some (or all) of the spread is due to problems with
the atmospheric models.

\item
Some of our spectra show signs of atmospheric inhomogeneities (star
spots), namely filling or reversals in Ca and Fe lines. We considered the
effects of star spots on the Li abundances, by using a two-component
synthesis, and found that the presence of such spots could explain the
Li abundance discrepancy found between the resonance and subordinate
lines. Differences in their depth of formation and excitation
potentials mean they react differently to changes in
temperature. Although there is some evidence that the most magnetically
active stars in our sample are indeed those that show significant
discrepancies in their Li abundances as estimated from the two lines;
counter examples are also seen.  Therefore, introduction of cooler regions on to the star
should only be regarded as a plausible explanation for
our observations at this stage.  Whilst star spots might be able to bring
abundances from the two lines into agreement, they cannot significantly
reduce the apparent spread in Li abundances among these cool Pleiades
objects.
\end{itemize}

\begin{acknowledgements}
The William Herschel and Isaac Newton Telescopes are operated on the
island of La Palma by the Isaac Newton Group in the Spanish
Observatorio del Roque de los Muchachos of the Instituto de Astrofisica
de Canarias. The authors acknowledge the travel and subsistence support
of the UK Particle Physics and Astronomy Research Council (PPARC). AF
was funded by a PPARC postgraduate studentship. Computational work was
performed on the Keele, St. Andrews and Open University nodes of the PPARC funded
Starlink network. This research has made extensive use of NASA's
Astrophysics Data System Abstract Service.
\end{acknowledgements}

\end{document}